\documentclass[12pt]{article}

\usepackage{newtxtext,newtxmath}

\usepackage{graphicx}
\usepackage{setspace}

\usepackage[letterpaper,margin=1in]{geometry}

\renewenvironment{abstract}
	{\quotation}
	{\endquotation}

\date{}

\makeatletter
\renewcommand{\fnum@figure}{\textbf{Figure \thefigure}}
\renewcommand{\fnum@table}{\textbf{Table \thetable}}
\makeatother

\usepackage{scicite}

\usepackage{url}

\usepackage{times}
\usepackage{amsmath,mathtools,nicematrix}
\usepackage{bbold}
\usepackage{bm}

\def\scititle{
	A Novel Solution of the Quantum Measurement Problem
}
\title{\bfseries \boldmath \scititle}

\author{
    Mani L. Bhaumik$^{1\ast}$\and
	\small$^{1}$Department of Physics and Astronomy, University of California, Los Angeles, USA 90095\and
	\small$^\ast$Email: bhaumik@physics.ucla.edu\and
}

\begin{document} 

\maketitle

\begin{abstract} 
\begin{spacing}{1}
{\noindent\textbf{Abstract.} In spite of the innumerable attempts to resolve the quantum measurement problem, almost since its beginning a century ago, a satisfactory solution still remains elusive. However, after the advent of quantum entanglement leading to environmental decoherence studies, a substantial leap forward has been achieved by Zurek and his associates. Their investigation has led to the existence of sturdy pointer states that survive decoherence leading to classicality although still remaining entangled with the environment and the apparatus states. But in order to avoid the inexplicable collapse postulate, Zurek invokes Everett's many world conjecture, which is not accepted by many experts. In this article, we present the observation of entanglement sudden death (ESD), demonstrating that disentanglement of the pointer states is caused by the quantum fluctuations of the electromagnetic field. We reveal that the decoherence between the pointer states and the environment is consistent with unitary evolution when the vacuum modes are included in the calculation, in analogy to the unitary evolution of the whole system in the spontaneous emission process. However, the ultimate state of the system is still subject to the totally unpredictable nature of individual quantum fluctuations, causing the stochastic appearance of an arbitrary classical state in von Neumann's collapse axiom even when the entire system evolves unitarily. Therefore, we suggest that an acceptable solution of the long standing measurement problem has finally been achieved.}
\end{spacing}
\end{abstract}

\newpage
\noindent
\section{Introduction}
In this International Year of Quantum Science and Technology, together with the celebration of the Centennial of nonrelativistic quantum mechanics, it is reasonable to consider quantum mechanics to be perhaps the most successful theory in modern physics with its broad applications. However, despite its spectacular success in predicting the results of experiments with uncanny accuracy, a satisfactory resolution of the quantum measurement problem has eluded us almost since its inception. 

The measurement problem arises from the fact that after a unitary evolution of a collection of quantum mechanically superposed states following the linear Schrödinger equation, the quantum superpositions upon measurement stochastically collapse into a definite classical outcome. Since John von Neumann proved that these processes are incompatible~\cite{neumann1932,neumann1955}, the von Neumann collapse postulate has been used as an axiom. The enigma of this axiom consists of several distinct questions:
(a) What causes the linear quantum evolution of the system to switch inexplicably from its microscopic unitary evolution suddenly to a non-unitary process with an unpredictable classical outcome including a certain probability? (b) Where do the probabilities come from? (c) What causes the wavefunction to collapse to a single member of the superposed state?

Over the years, an overabundance of solutions has been offered to explain this enigma. Unfortunately, so far none has been accepted unanimously without a reservation~\cite[p.~88]{weinberglec2013}. Of particular significance are the works by three dedicated groups with their devotion to the quantum measurement problem over a span of more than four decades. These are known as the Bohmian mechanics, GRW objective collapse theory and Zurek’s decoherence theory.

\subsection{Bohmian mechanics}
The earliest group since 1952, led by David Bohm and subsequently along with Basil Hiley~\cite{Bohm:1951xx,Bohm:1984,Bohm:1993}, postulates that a quantum particle like an electron is always a point particle and not represented by a wave function, but a wave of quantum potential or a pilot wave, guides the electron’s motion. This is in distinct contradiction to the latest research on a loophole-free derivation of the real wavefunction of a nonrelativistic particle like an electron~\cite{bhaumik2024}, showing that there is no real point particle before measurement. 

Furthermore, the so-called Bohmian mechanics postulates the existence of non-local hidden variables with no proof of their existence. Yet it posits that hidden variables cause an electron to always have a definite position even in the microscopic quantum domain, in contradiction to the presentation in~\cite{bhaumik2024}. Persuasive experimental demonstration of the phenomenon of entanglement, which garnered a Nobel prize~\cite{nobel2022}, removes any possibility of the existence of local hidden variables~\cite{Bell_1990}, putting the Bohmian mechanics out of the mainstream.

\subsection{GRW objective collapse theory}
The GRW model comprising the contributions of Ghirardi, Rimini, and Weber~\cite{GRW:1986} and its later version of the continuous spontaneous localization (CSL) model proposed by Philip Pearle~\cite{pearle:1989} along with its modification by Ghirardi, Pearle, and Rimini~\cite{Ghirardi:1990} do not involve participation of an observer. Although this is a very desirable property, unfortunately, the model requires a modification of the linear 
Schr\"{o}dinger equation with a nonlinear term. According to Steven Weinberg~\cite{weinberg:2016,weinberg:2016b,weinberg:2018}, the linearity of the Schr\"{o}dinger equation holds true with the accuracy of an atomic clock which puts a big constraint on modifying the equation. There are also several additional undesirable aspects, like the tail problem and position-only results, that are identified, among others, in the book “Decoherence” by Maximilian Schlosshauer~\cite{schlosshauer:2007a} and review article~\cite{Schlosshauer:2003zy}.

\subsection{Zurek's decoherence theory}
In an earlier article by the author~\cite{Bhaumik:2022ahb},
the importance of decoherence theory along with the recent discovery of the phenomenon of entanglement sudden death (ESD)~\cite{Yu_2004,Yu_2006a,Yu_2006b,Santos_2006,Cunha_2007,Yu_2007,Almeida_2007,ficek,lopez,Aolita_2008,PhysRevA.78.022322,Man_2009,Yu_2009} was pointed out to be perhaps the most promising approach in solving the measurement problem.

In the detailed studies of ESD presented in the above references, the effect of electromagnetic field quantum fluctuations was identified by several authors to be the cause of disentanglement. In the case of spontaneous emission, we have definite proof of the effect of quantum fluctuations on atomic states. In the anomalous $g$-factor of the spin magnetic moment, the effect of vacuum fluctuations is observed to be its primary cause~\cite{schwinger,fan_2023}, demonstrating that vacuum fluctuations affect not just atomic states. Since the non-local coherence of entanglement is also a property of the atomic states that can be quantified, distilled, and teleported, it is a relational feature between atoms arising from their shared states. In that sense, entanglement can be justifiably called some property of atoms. Therefore, it shouldn't be surprising to prudently expect that quantum fluctuations of the electromagnetic field will affect entanglement just as in the case of atomic and spin states. 

Decoherence is also the latest attempt to thoroughly investigate the measurement problem through many decades of dedicated work by competent groups of scientists. 
The idea of decoherence arose only after the discovery of the phenomenon of entanglement demonstrated by Alan Aspect, John F. Clauser, and Anton Zeilinger who were awarded the physics Nobel prize in 2022~\cite{nobel2022}. After the advent of quantum entanglement, Heinz-Dieter Zeh was the first to point out~\cite{Zeh:1970zz,Zeh_1973,Joos:1984uk} its utility in the decoherence process. When a quantum system containing several superposed states is exposed to the environment and the detector for the purpose of measurement, the linear Schrödinger evolution very quickly entangles the states of the quantum system ($|s_n\rangle$), environment ($|e_n\rangle$), and the detector ($|a_n\rangle$) into a product state 
$|s_n\rangle\otimes|a_n\rangle\otimes|e_n\rangle $
in the Hilbert space. This results in decoherence of the superposed states of the quantum system. 

Zeh got side tracked with his passion on Everette's many worlds interpretation (MWI) where the wavefunction continues to split indefinitely for every new observer, and the study of decoherence was quickly picked up by Wojciech Zurek and his group for its study over a long span of four decades. Zurek's studies came up with the existence of robust ``pointer" states representing classical states of an observable of the quantum states~\cite{Zurek:1982a,Paz,Zurek:2003a}. But the pointer states were still entangled with the environment and the detector.

Instead of using the usual state reduction formalism, Zurek, like Zeh, adopted MWI~\cite{Everett:1983,Everett:1983inbook} as well. Although it remains popular with some experts like S. Carroll~\cite{Sebens:2014iwa}, others see many problems with MWI. To begin with, MWI needs an observer, which contradicts the known absence of any conscious observer for quite some time in the early universe~\cite{Bell_1990}, yet definite indication of the quantum-to-classical transition exists~\cite{KIEFER_1998}. Of equal importance is the question posed by A. Kent~\cite{KENT_1990} and H. Stapp~\cite{Stapp_2002}, among others, where do probabilities arise in the MWI picture? 

The continual splitting of the wavefunction was also distasteful to renowned experts like Steven Weinberg~\cite{weinberg:2018,weinberglec2013},
who considered the collapse postulate equally unsatisfactory. Weinberg~\cite{weinberg:2018,weinberglec2013}, who devoted quite a bit of time on the subtleties of the nonrelativistic quantum mechanics, lamented quite recently the absence of a satisfactory resolution of the measurement problem despite many attempts over a century. 

\section{A novel concept}
In this article, we present an adaptation of the familiar phenomenon of spontaneous emission that has been around since the early days of quantum mechanics. The stocahstic nature of spontaneous emission appears to be quite similar to the stochastic nature of wavefunction collapse in quantum measurement. This similarity was already pointed out in an earlier article by the author~\cite{Bhaumik:2022ahb}.

It is well known that after the unitary evolution of a quantum state following the linear Schrödinger equation, a spontaneous emission of a photon from an upper quantum level of an atom appears stochastic. However, when the totality of the plethora of vacuum modes in all possible directions is included, the combined state of the atom with the bath of the universal fluctuations of the electromagnetic field undergoes unitary evolution. Nevertheless, the energy of the excited state is emitted unpredictably in a random direction. Thus the well-known process of spontaneous emission appears to be a stochastic process because of the inherent, infinitely random nature of the quantum fluctuations. 

In view of the recent discovery of the phenomenon of ESD, a similar situation appears to occur that could be used for the yet-to-be resolved quantum measurement problem. Again, after the relatively recent discovery of quantum entanglement and its application to decoherence in the quantum measurement process by Zurek and his coworkers, delineated in section \ref{sec:ZurekDecoherence}, an ensemble of entangled pointer states arises. As stated earlier, like Zeh, Zurek adapted the MWI approach as well.

Alternatively, the disentanglement of the pointer states leading to acquiring the value of an observable is accomplished by the utilization of the mysterious wavefunction collapse postulate laid down by John von Neumann in 1932. Discovery of the phenomenon of ESD caused by quantum fluctuations of the electromagnetic field suggests itself to determine its utilization in investigating the stochastic collapse of the entangled wavefunction suddenly and arbitrarily to a particular member of the entangled states. On careful examination, there seems to be a close similarity of the stochastic appearance of the spontaneous emission with the sudden emergence of an unpredictable quantum state even though the process involved in both events is unitary when all the vacuum modes are considered. A detailed mathematical analysis will now be presented in support of the observed similarity.

\section{Spontaneous emission}
Spontaneous emission is the process where an excited atom in the vacuum decays to the ground state by emitting a photon. This system can be analyzed in two different, yet equivalent, ways. First, the atom can be described as interacting with known dynamical degrees of freedom present in the vacuum, namely the vacuum fluctuations of the electromagnetic field, with the state of the whole system (atom and field) undergoing unitary evolution. 
Equivalently, we can consider the density matrix formulation, in which the vacuum electromagnetic field is traced out. The resulting reduced density matrix of the atom then undergoes non-unitary, stochastic evolution. We emphasize that the total density matrix of the entire system still evolves unitarily. Both models are described in the following sections.

\subsection{Interaction of atom with quantized electromagnetic field}
In the formalism outlined by Scully and Zubairy \cite{Scully:1997hcl}, the effect of the electromagnetic field on the two-state system of an atom is described. In this framework, an atom interacts with the quantized electromagnetic field. 
Specifically, the system is comprised of a two-level atom and an electromagnetic field decomposed into all possible modes of the photon field. The total Hamiltonian of the system is expressed as 
\begin{equation}
\begin{split}
\mathcal{H}&=\mathcal{H}_A+\mathcal{H}_F+\mathcal{H}_{\textrm{int}}
\end{split}
\end{equation}
Here, $\mathcal{H}_A$ represents the Hamiltonian of the free atom, $\mathcal{H}_F$ represents the Hamiltonian of the free electromagnetic field, and $\mathcal{H}_{\textrm{int}}$ represents the interaction of the atom and the electromagnetic field. The total system described by $\mathcal{H}$ is closed and therefore evolves unitarily, with its evolution described by the Schr\"{o}dinger equation.

The Hamiltonian of the atom is 
\begin{equation}
    \mathcal{H}_A=\frac{1}{2}\hbar \omega\sigma_z+\frac{1}{2}(E_a+E_b)
\end{equation}
The two possible states are the excited state $|a\rangle$ and the non-excited state $|b\rangle$, and the difference in energy is $E_a-E_b=\hbar\omega$, where $\omega$ is called the transition frequency.
The Pauli matrices $\sigma_\pm,\sigma_z$ are given by 
\begin{equation}
\begin{split}
    \sigma_-&=\begin{pmatrix}
0 & 0\\
1 & 0
\end{pmatrix}\\
\sigma_+&=\begin{pmatrix}
0 & 1\\
0 & 0
\end{pmatrix}\\
\sigma_z&=\begin{pmatrix}
1 & 0\\
0 & -1
\end{pmatrix}
\end{split}
\end{equation}

The Hamiltonian of the free electromagnetic field is written as an infinite sum over all photon modes, modeled as individual harmonic oscillators
\begin{equation}
    \mathcal{H}_F=\sum_{\bm{k}}\hbar\nu_k\left(a_{\bm{k}}^\dag a_{\bm{k}}+\frac{1}{2}\right)
\end{equation}
The operators $a_{\bm{k}}^\dag$ and $a_{\bm{k}}$ are the creation and annihilation operators, respectively, of the photon with wavevector $\bm{k}$ and frequency $\nu_k=c |\bm{k}|$.

The interaction is modeled using the dipole approximation and the rotating-wave approximation and is given by 
\begin{equation}
    \mathcal{H}_{\textrm{int}}=\hbar\sum_{\bm{k}}g_{\bm{k}}\left(\sigma_+ a_{\bm{k}}+a_{\bm{k}}^\dag \sigma_-\right)
\end{equation}
The coupling constant $g_{\bm{k}}$ describes the strength of the electric-dipole interaction, mediating the interaction between the atom and the electromagnetic field. We now calculate the evolution of the state of the system under this interaction Hamiltonian, following the Weisskopf-Wigner model.

In the Weisskopf-Wigner model \cite{Scully:1997hcl} of spontaneous emission, the time evolution of the observables in the interaction picture is determined by the free part of the total Hamiltonian $\mathcal{H}_A+\mathcal{H}_F$ and the time evolution of the states is determined by the interaction $\mathcal{H}_{\textrm{int}}$.
Therefore, the interaction-picture Hamiltonian is 
\begin{equation}
\begin{split}
\mathcal{V}&=e^{i(\mathcal{H}_A+\mathcal{H}_F)t/\hbar}\mathcal{H}_{\textrm{int}}e^{-i(\mathcal{H}_A+\mathcal{H}_F)t/\hbar}\\
&=\hbar\sum_{\bm{k}}\left(g_{\bm{k}}^*e^{i\bm{k}\cdot\bm{r}}\sigma_+a_{\bm{k}}e^{i(\omega-\nu_k)t}+\textrm{Hermitian conjugate}\right)
\end{split}
\end{equation}
where $\bm{r}$ is the position of the atom.

At $t=0$, the combined system is assumed to be initially in the state $|a,0\rangle$, where the atom is in the excited state $|a\rangle$ and the electromagnetic field is in the vacuum state $|0\rangle$. The state of the system at time $t$ is then:
\begin{equation}
    |\psi(t)\rangle=c_a(t)|a,0\rangle+\sum_{\bm{k}}c_{b,\bm{k}}(t)|b,1_{\bm{k}}\rangle
\end{equation}
with initial conditions $c_a(0)=1$ and $c_{b,\bm{k}}(0)=0$. The second term is the sum over all possible configurations where the atom is in the ground state and one photon has been emitted with wavevector $\bm{k}$.

Substituting this state into the Schrodinger equation
\begin{equation}
\frac{d|\psi(t)\rangle}{dt}=-\frac{i}{\hbar}\mathcal{V}|\psi(t)\rangle
\end{equation}
results in a set of coupled first-order differential equations for the coefficients $c_a(t)$ and $c_{b,\bm{k}}(t)$. To solve this system, the field mode frequencies $\nu_k$ are approximated as a continuum and it is assumed that the intensity of the emitted radiation is approximately equal to $\omega$. Therefore, the solution is
\begin{equation}
\begin{split}
    c_a(t)&=e^{-\Gamma t/2}\\
    c_{b,\bm{k}}(t)&=g_{\bm{k}}e^{-i\bm{k}\cdot\bm{r}}\left(\frac{1-e^{-i(\omega-\nu_k)t-\Gamma t/2}}{\nu_k-\omega+i\Gamma/2}\right)
\end{split}
\end{equation}
The decay constant $\Gamma$ of the atom in the excited state is given by
\begin{equation}\label{eq:decayrate}
    \Gamma=\frac{1}{4\pi\epsilon_0}\frac{4\omega^3(e\langle a|\bm{r}|b\rangle)^2}{3\hbar c^3}
\end{equation}
where $e\langle a|\bm{r}|b\rangle$ is the electric-dipole transition matrix element of the atom at the position $\bm{r}$. As usual, $e$ is the electric charge and $\epsilon_0$ is the vacuum permittivity.

It is important to note that without considering the interaction of the excited atom with the vacuum electromagnetic field, spontaneous emission appears to be a stochastic process. However, when the interaction with the vacuum is included, the evolution of the atom into the state $|\psi(t)\rangle$ is a unitary process. The decay rate, for example, is computed from the unitary evolution of the combined atom-electromagnetic field system. For further clarification, although the stable expectation value of the vacuum modes is included in the calculations, there are an infinite number of fluctuations which occur in all possible directions and each one is totally random. 
The final particular excited mode of the electromagnetic field, emitted from the atom into the vacuum, is therefore unpredictable.

\subsection{Density matrix formalism}
Now the above treatment of spontaneous emission will be explored using the density matrix formalism, especially since it allows for the treatment of both pure and mixed quantum states, as well as the inclusion of decoherence and dissipation. 
This formalism is presented here for direct comparison with the process of decoherence, which is particularly in terms of density matrices. This additionally highlights the similarities between spontaneous emission and decoherence in the context of the quantum measurement problem.

In the two-level atom model (with excited state $|a\rangle$ and ground state $|b\rangle$), spontaneous emission corresponds to a process where the system decays from $|a\rangle$ to $|b\rangle$ by emitting a photon. The reduced density matrix of the atom is 
\begin{equation}
    \rho=\begin{pmatrix}
\rho_{aa} & \rho_{ab} \\
\rho_{ab}^* & \rho_{bb}
\end{pmatrix}
\end{equation}
Here, the degrees of freedom of the vacuum electromagnetic field are traced out. Therefore, the evolution of the reduced density matrix of the atom cannot be described by the Schrödinger equation alone. Instead, there is a master equation describing its evolution, typically given in Lindblad form:
\begin{equation}
\frac{d\rho}{dt} = -\frac{i}{\hbar}[H, \rho] + \mathcal{L}[\rho]
\end{equation}
where $H$ is the Hamiltonian of the atom and $\mathcal{L}[\rho]$ is the Lindblad dissipator that models spontaneous emission.

For spontaneous emission (with rate $\Gamma$), the dissipator is
\begin{equation}
\mathcal{L}[\rho] = \Gamma \left( \sigma_- \rho \sigma_+ - \frac{1}{2} \{\sigma_+ \sigma_-, \rho\} \right)
\end{equation}
where the Pauli matrices are $\sigma_- = |b\rangle\langle a|$ (lowering operator), $\sigma_+ = |a\rangle\langle b|$ (raising operator), and the curly brackets $\{A, B\}$ denote the anticommutator between two operators $A$ and $B$.
As shown in chapter 8 of \cite{Scully:1997hcl}, this term causes the population to decay from the excited state to the ground state, and coherence (off-diagonal elements) to decay due to decoherence. The interpretation of this process is that the excited state population (density matrix element $\rho_{aa}$) decays at rate $\Gamma$, and the ground state population ($\rho_{bb}$) increases accordingly. Additionally, the statement of coherence loss is that the off-diagonal matrix element $\rho_{ab}$ decays at a rate $\Gamma/2$ due to decoherence from spontaneous emission. The authors of \cite{Scully:1997hcl} emphasize that the atomic decay rate derived from this model (with appropriate assumptions) is equivalent to the result eq.~\eqref{eq:decayrate}.

In summary, spontaneous emission is naturally and effectively described in the density matrix formalism via the Lindblad master equation. This approach is convenient when considering open quantum systems and is widely used in quantum optics and quantum information theory.

\section{Decoherence}\label{sec:ZurekDecoherence}
Decoherence is the process by which a quantum system loses its coherent superposition states due to interactions with its environment, leading to the emergence of classical behavior. The theory of decoherence is described by Zurek and further explained by Schlosshauer \cite{Zurek:2003zz,Schlosshauer:2003zy}. 
To derive decoherence, consider a closed system comprised of the quantum system of interest $\mathcal{S}$, a measurement apparatus $\mathcal{A}$, and the environment $\mathcal{E}$. Each has its own states that make up its Hilbert space:
\begin{equation}
    |s_n\rangle\in \mathcal{H}_{\mathcal{S}}\,,\quad |a_n\rangle\in \mathcal{H}_{\mathcal{A}}\,,\quad |e_n\rangle\in \mathcal{H}_{\mathcal{E}}
\end{equation}
where $|s_n\rangle$ is a possible state of the system $\mathcal{S}$ and $\mathcal{H}_{\mathcal{S}}$ is its Hilbert space, and similarly for the apparatus and environment. 

The state of the combined system initially evolves during the ``premeasurement" process into
\begin{equation}
    |\psi\rangle = \sum_n c_n |s_n\rangle|a_n\rangle|e_n\rangle
\end{equation}
where $c_n$ are complex coefficients, $|a_n\rangle$ are the so-called \textit{pointer states} of the apparatus, and $|e_n\rangle$ are the corresponding pointer states of the environment.
The combined $\mathcal{S}\mathcal{A}\mathcal{E}$ system in this model evolves unitarily since it is a closed system.

The experimentally-accessible subsystem of the system-apparatus $\mathcal{S}\mathcal{A}$ is described using the reduced density matrix $\rho_{\mathcal{S}\mathcal{A}}$. $\rho_{\mathcal{S}\mathcal{A}}$ is obtained by tracing out the degrees of freedom of the environment from the total density matrix $\rho_{\mathcal{S}\mathcal{A}\mathcal{E}}$:
\begin{equation}
    \rho_{\mathcal{S}\mathcal{A}}=\textrm{Tr}_{\mathcal{E}}\left(\rho_{\mathcal{S}\mathcal{A}\mathcal{E}}\right)=\sum_{k,n} c_k c_n^* |s_k\rangle|a_k\rangle \langle s_n|\langle a_n| \langle e_n|e_k\rangle 
\end{equation}
Interference terms between different pointer states are present if $\langle e_n|e_k\rangle\neq 0$ for $k\neq n$. 

Many explicit models of the system-environment interactions show that over time, $\langle e_n|e_k\rangle\rightarrow 0$ for $k\neq n$. This is the essence of decoherence: due to the large number of degrees of freedom in the environment, local coherence between pointer states in the reduced $\mathcal{S}\mathcal{A}$ subsystem is lost. In the pointer basis, the reduced density matrix becomes 
\begin{equation}
    \rho_{\mathcal{S}\mathcal{A}}=\sum_n |c_n|^2 |s_n\rangle|a_n\rangle \langle s_n|\langle a_n|
\end{equation}
In the pointer basis, the reduced density matrix is diagonal, which signifies that the local interference terms have vanished and local coherence has been lost, hence the phrase ``decoherence."
Explicit calculations modeling the system-environment interaction typically involve treating the environment as a collection of harmonic oscillators \cite{Caldeira:1982iu,Hu:1991di,Joos2003-JOOEDA,PhysRevD.40.1071,PhysRevLett.70.1187,Zurek:2003zz}. 
It is important to note that the combined $\mathcal{S}\mathcal{A}\mathcal{E}$ system described by $\rho_{\mathcal{S}\mathcal{A}\mathcal{E}}$ retains non-local coherence. The pointer states of the apparatus and system are entangled with the states of the environment.

\section{Decoherence with the vacuum}
Building on Zurek’s formulation \cite{Zurek:2003zz}, the model is extended to include the degrees of freedom of the quantum fields present in the vacuum~\cite{Bhaumik:2023ugp}. The interaction of the environment with the vacuum electromagnetic field is considered. Let $\mathcal{V}$ represent the subsystem corresponding to the electromagnetic field, with its states and Hilbert space given by
\begin{equation}
    |v_n\rangle \in \mathcal{H}_\mathcal{V}
\end{equation}
The total system $\mathcal{S}\mathcal{A}\mathcal{E}\mathcal{V}$ is a closed system and evolves unitarily. After the premeasurement, the electromagnetic field couples to the states in the $\mathcal{S}\mathcal{A}\mathcal{E}$ subsystem, so the state of the entire system $|\widetilde{\psi}\rangle$ is
\begin{equation}
    |\widetilde{\psi}\rangle = \sum_n \widetilde{c}_n |s_n\rangle|a_n\rangle|e_n\rangle|v_n\rangle
\end{equation}
for some complex coefficients $\widetilde{c}_n$. 

Upon tracing out the degrees of freedom associated with the electromagnetic field, the $\mathcal{S}\mathcal{A}\mathcal{E}$ subsystem becomes an \textit{open} system and the reduced density matrix is 
\begin{equation}
\rho_{\mathcal{S}\mathcal{A}\mathcal{E}}=\textrm{Tr}_{\mathcal{V}}\left(\rho_{\mathcal{S}\mathcal{A}\mathcal{E}\mathcal{V}}\right)=\sum_{k,n} \widetilde{c}_k \widetilde{c}_n^{\,*} |s_k\rangle|a_k\rangle |e_k\rangle \langle s_n|\langle a_n|\langle e_n| \langle v_n|v_k\rangle
\end{equation}
In this case, there are interference terms present in the reduced density matrix due to $\langle v_n|v_k\rangle$ for $n\neq k$. 
Fluctuations in the electromagnetic field lead to amplitude noise in the reduced density matrix, which destroys the interference terms, causing decoherence at the level of the reduced density matrix as will be shown below.

For convenience, the experimentally-accessible subsystem $\mathcal{S}\mathcal{A}$ is denoted as the subsystem $\mathcal{M}$, with pointer states
\begin{equation}
     |m_n\rangle\equiv|s_n\rangle|a_n\rangle 
\end{equation}
existing in the Hilbert space 
\begin{equation}
    \mathcal{H}_{\mathcal{M}}\equiv\mathcal{H}_{\mathcal{S}} \otimes \mathcal{H}_{\mathcal{A}}
\end{equation}
The reduced density matrix of the system-apparatus-environment subsystem is then written as
\begin{equation}
\rho_{\mathcal{M}\mathcal{E}}=\sum_{k,n} \widetilde{c}_k \widetilde{c}_n^{\,*} |m_k\rangle |e_k\rangle \langle m_n|\langle e_n| \langle v_n|v_k\rangle
\end{equation}

The states of the electromagnetic field are modeled as a collection of an infinite number of harmonic oscillators. The measurement-environment subsystem, $\mathcal{M}\mathcal{E}$, is treated as a collection of two-state systems called qubits. Specifically, the $\mathcal{M}\mathcal{E}$ subsystem is modeled as two qubits that are initially entangled. There may be a large number of qubits in the $\mathcal{M}\mathcal{E}$ subsystem, but it is assumed for our calculation that they are only pairwise entangled. The two-level qubit approximation serves as an intuitive model for the correlations between the pointer states of the measurement and the environment subsystems.

The measurement-environment subsystem in this model is described by a two-qubit open quantum system, interacting with a bath of quantum harmonic oscillators, that is, the electromagnetic field. As shown in \cite{Yu_2009}, the reduced density matrix for the $\mathcal{M}\mathcal{E}$ subsystem, modeled by two qubits, can be written
\begin{equation}
    \rho_{\mathcal{M}\mathcal{E}}=\begin{pmatrix}
a & 0 & 0 & y\\
0 & b & z & 0\\
0 & z^* & c & 0\\
y^* & 0 & 0 & d
\end{pmatrix}
\end{equation}
where $a,b,c,d,y,z$ are arbitrary complex numbers and $d=1-a$. The matrix is written in the basis $\{|1,1\rangle,|1,0\rangle,|0,1\rangle,|0,0\rangle\}$ where the possible states of each qubit are $|0\rangle$ and $|1\rangle$. In this pointer basis, decoherence means the off-diagonal elements go to zero, that is, $y,z\rightarrow 0$. 

The fluctuations of the electromagnetic field, present even in the vacuum, 
introduce amplitude noise that dampens the interference terms of the open quantum system $\mathcal{M}\mathcal{E}$. Again, one could use the Lindblad master equation to describe the evolution of the density matrix in the presence of noise, but the Lindblad equation is complicated and difficult to deal with. Instead, the noise can be modeled using a set of much simpler quantum operators called Kraus operators \cite{Nielsen_Chuang_2010, Yu_2009}. These operators $K_n(t)$ describe the time evolution of a density matrix (initially $\rho(0)$) for an open system:
\begin{equation}\label{eq:evolution}
    \rho(t)=\sum_n K_n(t)\rho(0)K_n^\dag (t)
    \end{equation}
with the Kraus operators satisfying
    \begin{equation}
        \sum_n K_n^\dag (t)K_n(t)=\mathbb{1}
\end{equation}

For amplitude noise, such as that induced by vacuum fluctuations, the Kraus operators are given by \cite{Yu_2009}
\begin{NiceMatrixBlock}[auto-columns-width]
    \begin{align}
        K_1(t)&=\begin{pNiceMatrix}
            \gamma(t)^2 & 0 & 0 & 0\\
            0 & \gamma(t) & 0 & 0\\
            0 & 0 & \gamma(t) & 0\\
            0 & 0 & 0 & 1
        \end{pNiceMatrix}\\
        K_2(t)&=\begin{pNiceMatrix}
            0 & 0 & 0 & 0\\
            \gamma(t)\omega(t) & 0 & 0 & 0\\
            0 & 0 & 0 & 0\\
            0 & 0 &\omega(t) & 0
        \end{pNiceMatrix}\\
        K_3(t)&=\begin{pNiceMatrix}
            0 & 0 & 0 & 0\\
            0 & 0 & 0 & 0\\
            \gamma(t)\omega(t) & 0 & 0 & 0\\
            0 & \omega(t) & 0 & 0
        \end{pNiceMatrix}\\
        K_4(t)&=\begin{pNiceMatrix}
            0 & 0 & 0 & 0\\
            0 & 0 & 0 & 0\\
            0 & 0 & 0 & 0\\
            \omega(t)^2 & 0 & 0 & 0
        \end{pNiceMatrix}\\
        \nonumber
    \end{align}
\end{NiceMatrixBlock}
where
\begin{equation}
    \gamma(t)=e^{-\Gamma t/2},\qquad\omega(t)=\sqrt{1-\gamma(t)^2}
\end{equation}
and $\Gamma>0$ represents the strength of the amplitude noise.
After incorporating the Kraus operators into equation \eqref{eq:evolution}, the result for the off-diagonal elements decays with time:
\begin{equation}
    \begin{split}
        y(t)&=\gamma(t)^2y\rightarrow 0\textrm{ as }t\rightarrow \infty\\
        z(t)&=\gamma(t)^2z\rightarrow 0\textrm{ as }t\rightarrow \infty
    \end{split}
\end{equation}
for arbitrary constants $y,z$.
Therefore, after a time $t$ such that $y(t), z(t)$ are sufficiently small, the density matrix of the measurement-environment subsystem becomes 
\begin{equation}
    \rho_{\mathcal{M}\mathcal{E}}=\begin{pmatrix}
a(t) & 0 & 0 & 0\\
0 & b(t) & 0 & 0\\
0 & 0 & c(t) & 0\\
0 & 0 & 0 & d(t)
\end{pmatrix}
\end{equation}
where the functions $a(t),b(t),c(t),d(t)$ are calculated using the Kraus operators:
\begin{equation}
    \begin{split}
        a(t)&=a\, \gamma(t)^4\\
        b(t)&=\gamma(t)^2\left(b+a\,\omega(t)^2\right)\\
        c(t)&=\gamma(t)^2\left(c+a\,\omega(t)^2\right)\\
        d(t)&=1-a+(b+c)\omega(t)^2+a\,\omega(t)^4
    \end{split}
\end{equation}

This calculation shows that for an open two qubit system, modeling a quantum system and measurement apparatus interacting with an environment, which is affected by amplitude noise such as that induced by vacuum fluctuations, coherence is exponentially suppressed. It should be emphasized that although any one of these subsystems is considered non-unitary, the entire combined ensemble is a closed system and therefore evolves unitarily.

By including the effect of the vacuum modes on the quantum system, we have successfully shown that the non-local coherence of Zurek's pointer states and the environment is destroyed. The distribution of the final observed state of the system is given by the reduced density matrix $\rho_{\mathcal{M}\mathcal{E}}$, and we contend that the final observed state is determined stochastically from this distribution through the influence of vacuum quantum fluctuations. For further clarification, we would like to recall F. Wilczek’s remark~\cite{wilczek:2006a}, “loosely speaking, energy can be borrowed to make evanescent virtual particles. Each pair passes away soon after it comes into being, but new pairs are constantly boiling up, to establish an equilibrium distribution.” In spite of the equilibrium distribution, there are individual fluctuations that are totally unpredictable.

\section{Concluding Remarks}
Perhaps the most significant result produced by Zurek and his coworkers in their decades long research on decoherence is the existence of pointer states. These states arise when a quantum system of superposed states is exposed to the environment and the measuring apparatus for the purpose of carrying out quantum measurement. In  a very short time, evolution described by the Schrödinger equation entangles the quantum system with the environment and the apparatus states and the entire system evolves unitarily.

Zurek showed that the sturdy pointer states corresponding to the classical outcomes remain stable during decoherence of the local coherence of the superposed states. However, the pointer states remain entangled in a non-local coherence with the environment as well as the measuring apparatus. Zurek then used the many worlds interpretation to get the unentangled classical observables.

Earlier, we have pointed out some of the undesirable features of the many worlds interpretation. We contend that the pointer states can be disentangled by the action of the universal quantum fluctuations of the electromagnetic field as recently demonstrated by the phenomenon of entanglement sudden death in bipartite entanglement. In this paper, we considered only pairwise interactions between qubits, that is, bipartite entanglement. However, there is reason to believe that bipartite entanglement can be extended to multipartite entanglement of pointer states~\cite{Zhang_2010,Cole_2010}, since their entanglement exhibits greater fragility, similar to that of GHZ states, in contrast to the comparatively robust nature of W states.

Thus, the disentanglement of the pointer states caused by the universal quantum fluctuations appears quite like the familiar spontaneous emission. While both cases look stochastic, in fact the total systems evolve unitarily. In both cases the stochastic nature arises due to the completely spontaneous and unpredictable characteristics of individual quantum fluctuations giving rise to a vacuum mode~\cite{wilczek:2006a}. Therefore, the sudden, inexplicable appearance of a particular state to  which the wave function collapses in the axiom of von Neumann’s collapse postulate can be reasonably explained to be stochastic while the total system evolves unitarily.

At this juncture, it will be important to point out a rather ironic aspect of our presentation. While the disentanglement caused by quantum fluctuations is essential for applications in quantum measurement, it is the very phenomena that is totally undesirable for the extensive work being carried out in the development of quantum computers. This situation is resolved by the application of error code in the early stages of disentanglement. Studies indeed indicate that ESD can be prevented by applying error correction type measures in the early stages of the process~\cite{Xie:2022ozf}.

In a previous work by the author~\cite{Bhaumik:2021yku}, a cogent explanation was provided for the emergence of the familiar Born rule. However, this leaves open the fundamental question of how the entire wave function collapses to a single observed state. In addressing this issue, it is important to recall that a quantum of an observable—represented by an ensemble of superposed states—is inherently ``holistic,'' such that the components ``are parts of the one entangled whole"~\cite{penrose}. This naturally leads to the question: where does the remainder of the quantum of the observable go, given that the eigenvalue of the reduced state reflects only a portion of the whole quantum? A persuasive premise~\cite[p.~511]{penrose} is that the rest of the energy becomes inaccessible, effectively residing in the unobservable part of the fully entangled system, which includes the environment and the measurement apparatus.

\section*{Acknowledgments}
The author wishes to acknowledge very helpful discussions with Zvi Bern and Anna Wolz. 

\clearpage 

\bibliography{scibib} 
\bibliographystyle{sciencemag}

\end{document}